# QUANTISING GENERAL RELATIVITY USING QED THEORY

Sarah B. M. Bell[1,2] and Bernard M. Diaz[1,3]

## Abstract

We apply QED theory to quantum gravity and find it leads to general relativity in the classical limit. We discuss the implications of the result for the quantum-classical divide. This enables us to relate our result to M-theory.

## 1. GENERAL RELATIVITY AND QED THEORY

### 1.1 The equation of circular motion

We broaden our previous discussion[(5-8)] of QED theory by including general relativity. We introduce a space $L_1$ with a metric,

$$\mathrm{d}\hat{\tau}^2 = \mathrm{d}\hat{x}_0{}^2 - \mathrm{d}\hat{x}_1^2 - \mathrm{d}\hat{x}_2^2 - \mathrm{d}\hat{x}_3^2 \qquad (1.1.A)$$

where $(\hat{x}_0, \hat{x}_1, \hat{x}_2, \hat{x}_3)$ are Cartesian co-ordinates, with $\hat{x}_0$ temporal. We have set the speed of light, $c$, and the gravitational constant, $G$, to 1. Addressing the $(\hat{x}_1, \hat{x}_2)$ plane of $L_1$ using polar co-ordinates, $(\hat{r}, \hat{\theta})$, the metric becomes

[1] Department of Computer Science I.Q. Group, The University of Liverpool, Chadwick Building, Peach Street, Liverpool, L69 7ZF, United Kingdom.

[2] UK telephone number and email address, 01865 798579 and Sarabell@dial.pipex.com

[3] Email address, B.M.Diaz@csc.liv.ac.uk

$$\mathrm{d}\hat{\tau}^2 = \mathrm{d}\hat{x}_0^2 - \hat{r}^2\,\mathrm{d}\hat{\theta}^2 - \mathrm{d}\hat{r}^2 - \mathrm{d}\hat{x}_3^2 \tag{1.1.B}$$

We introduce a point source, of constant positive mass $m_s$, at the origin of a space $C_1$ with a metric,

$$\mathrm{d}\tau^2 = \frac{2m_s}{r}\mathrm{d}s^2 - \frac{r}{2m_s}\mathrm{d}r^2 \tag{1.1.C}$$

where *s* is the temporal and *r* a spatial co-ordinate. This is a solution of the Einstein equation,(19) although it is not unique.(25, 28) A point particle we call *the preon* moves in a circular orbit in $L_1$ and an orbit with $\mathrm{d}r = 0$ in $C_1$. We analyse the motion in each case and then relate our results.

We find the equation of circular motion in $L_1$.(25) The non-zero Christoffel symbols(28) for the metric in equation (1.1.B) are

$$\Gamma^{\hat{\theta}}{}_{\hat{r}\hat{\theta}} = \Gamma^{\hat{\theta}}{}_{\hat{\theta}\hat{r}} = \frac{1}{\hat{r}}, \quad \Gamma^{\hat{r}}{}_{\hat{\theta}\hat{\theta}} = -\hat{r} \tag{1.1.D}$$

The geodesic equation(25) for co-ordinates $x^\mu$ and interval $\mathrm{d}\tau$ is

$$\frac{\mathrm{d}^2 x^\mu}{\mathrm{d}\tau^2} + \Gamma^\mu_{\nu\rho}\frac{\mathrm{d}x^\nu}{\mathrm{d}\tau}\frac{\mathrm{d}x^\rho}{\mathrm{d}\tau} = 0 \tag{1.1.E}$$

which becomes

$$\frac{\mathrm{d}^2\hat{r}}{\mathrm{d}\hat{\tau}^2} = \hat{r}\left(\frac{\mathrm{d}\hat{\theta}}{\mathrm{d}\hat{\tau}}\right)^2 \tag{1.1.F}$$

from equations (1.1.D). If the velocity along the arc is $v_g$,

$$v_g = \hat{r}\frac{\mathrm{d}\hat{\theta}}{\mathrm{d}t}, \quad \frac{\mathrm{d}^2\hat{r}}{\mathrm{d}t^2} = \frac{v_g^2}{\hat{r}} \tag{1.1.G}$$

from the metric in equation (1.1.B) with $\mathrm{d}\hat{r} = \mathrm{d}\hat{x}_3 = 0$ and $t = \hat{x}_0$.

We find the equivalent in $C_1$. The non-zero Christoffel symbols for the metric in equation (1.1.C) are

$$\Gamma^r_{rr} = \frac{1}{2r}, \quad \Gamma^s_{rs} = \Gamma^s_{sr} = -\frac{1}{2r}, \quad \Gamma^r_{ss} = -\frac{2m_s^2}{r^3} \tag{1.1.H}$$

The geodesic equation (1.1.E) becomes

$$\frac{\mathrm{d}^2 r}{\mathrm{d}\tau^2} + \frac{1}{2r}\left(\frac{\mathrm{d}r}{\mathrm{d}\tau}\right)^2 - \frac{2m_s^2}{r^3}\left(\frac{\mathrm{d}s}{\mathrm{d}\tau}\right)^2 = 0 \tag{1.1.I}$$

Since,

$$\frac{\mathrm{d}s}{\mathrm{d}\tau} = \sqrt{\frac{r}{2m_s}} \tag{1.1.J}$$

from the metric in equation (1.1.C) with $\mathrm{d}r = 0$, we obtain,

$$\frac{\mathrm{d}^2 r}{\mathrm{d}\tau^2} = \frac{m_s}{r^2} \tag{1.1.K}$$

from equation (1.1.I).

## 1.2 The quantum condition

We describe the preon's motion by the parallel transport of a vector giving its position round the orbit.[(28)] For Cartesian co-ordinates in a flat space this is the same, except for the sign of the angle turned, as changing the orientation of the position vector with respect to fixed axes. We find the behaviour of the position vectors for $L_1$ and $C_1$ in $L'$ and $C'$, respectively, the flat two-dimensional spaces over which we suppose that the vectors are defined. The position vector in $L'$ is given by,

$$\mathbf{V}_{L'} = -\begin{pmatrix} \hat{x}_1 \\ \hat{x}_2 \end{pmatrix} \tag{1.2.A}$$

The angle through which the vector has turned in time *t* is

$$\hat{\theta} = \frac{v_g t}{\hat{r}} \tag{1.2.B}$$

from the integration of the first of equations (1.1.G).

We find the orbit in $C'$. If the position vector in $C'$ is given by,

$$\mathbf{V}_{C'} = \begin{pmatrix} V^s \\ V^r \end{pmatrix} \tag{1.2.C}$$

the equation for parallel transport[28] is

$$\frac{\mathrm{d}\mathbf{V}_{C'}}{\mathrm{d}\tau} = -\frac{\mathrm{d}x^{\sigma}}{\mathrm{d}\tau}\boldsymbol{\Gamma}_{\sigma}\mathbf{V}_{C'} \tag{1.2.D}$$

where the matrices $\boldsymbol{\Gamma}$ are assembled from the Christoffel symbols by the prescription that $\Gamma^{\alpha}_{\sigma\beta}$ is the element in row $\alpha$ and column $\beta$ of the matrix $\boldsymbol{\Gamma}_{\sigma}$. We obtain,

$$\boldsymbol{\Gamma}_s = \begin{pmatrix} 0 & -1/2r \\ -2m_s^2/r^3 & 0 \end{pmatrix} \tag{1.2.E}$$

from equations (1.1.H). Solving equation (1.2.D) using equations (1.1.J) and (1.2.C),

$$V^s = -\frac{irD}{2m_s}\sin\left(\frac{\tau}{r}\sqrt{-\frac{m_s}{2r}} + F\right), \qquad (1.2.F)$$
$$V^r = D\cos\left(\frac{\tau}{r}\sqrt{-\frac{m_s}{2r}} + F\right)$$

where $D$ is a real and $F$ an imaginary constant of integration. We move from the orbit in $C'$ to a circular one in a complex space, $C'^*$, as we did in similar cases previously,[5, 8] defining,

$$\mathbf{V}_{C'}{}^* = \begin{pmatrix} iV^s \\ \dfrac{r}{2m_s}V^r \end{pmatrix} \qquad (1.2.G)$$

The imaginary angle, $i\theta$, through which $\mathbf{V}_{C'}{}^*$ has turned in $C'^*$ after an interval $\tau$ is given by,

$$\theta = \frac{\tau}{r}\sqrt{\frac{m_s}{2r}} \qquad (1.2.H)$$

from equation (1.2.F) and where we have chosen a sign.

If we suppose the relationship between $L_1$ and $C_1$ is given by,

$$\hat{\theta} = \theta\sqrt{2} \qquad (1.2.I)$$

then equations (1.1.G), (1.1.K), (1.2.B) and (1.2.H) are consistent with the assignments,

$$t = \tau, \quad \hat{r} = r, \quad v_g = \sqrt{\frac{m_s}{\hat{r}}}, \quad \frac{\mathrm{d}^2\hat{r}}{\mathrm{d}t^2} = \frac{m_s}{\hat{r}^2} \qquad (1.2.J)$$

It will be convenient to describe the movement of the preon in $C_1$ as a rotation in what follows. We also omit the $\hat{}$ on $\hat{r}$ where it is identical to *r*, unless we wish to emphasise the distinction.

We define *versatile quantisation* or *v-quantisation*,

$$v_g = \frac{m}{n} \tag{1.2.K}$$

where *m* and *n* are integers. Equations (1.2.J) and (1.2.K) constrain the radius, *r*, to discrete values, *R*, which we call the *Bohr radii*, but they leave the system unconstrained in the limit when $m, n \to \infty$, which we interpret as the classical case. We associate a wave with the preon, which is only defined on the $(\hat{x}_1, \hat{x}_2)$ plane in $L_1$, so that the metric is

$$\mathrm{d}\tilde{\tau}^2 = r^2\,\mathrm{d}\hat{\xi}^2 + \mathrm{d}r^2 \tag{1.2.L}$$

for polar co-ordinates, $(r, \hat{\xi})$, where $\hat{\xi}$ is the phase of the wave. When $\mathrm{d}r = 0$, the phase change is

$$\hat{\xi} = \frac{\tilde{\tau}}{r} \tag{1.2.M}$$

from integrating the metric in equation (1.2.L). We eliminate *t* from equation (1.2.B) using equations (1.2.J),

$$\hat{\theta} = \frac{v_g \tau}{\hat{r}} \tag{1.2.N}$$

Equations (1.2.K), (1.2.M) and (1.2.N) with $r = R$ are consistent with,

$$\tilde{\tau} = m\tau, \quad \hat{\xi} = n\hat{\theta} \tag{1.2.O}$$

We suppose $\tilde{\tau}$ is the interval for the wave in $C_1$. From equation (1.2.H),

$$\xi = \frac{\tilde{\tau}}{R}\sqrt{\frac{m_s}{2R}}, \quad \xi = m\theta \tag{1.2.P}$$

where $\xi$ is the phase change in $C_1$. We call *n the excitation number* and *m the winding number* of the preon.

## 1.3 Electromagnetism and gravity compared

We described QED theory in the context of electromagnetism using an inverse distance-squared law for the force and *Bohr's equations* in the interrelated spaces *S*, *M* and *L*.(6-8) *S* and *M* map to $L$ and we use only $L$ where possible here. Bohr's equations for electromagnetism(6) are

$$\frac{e_e e_n}{R_e} = \frac{m_e v_e^2}{\sqrt{1 - v_e^2}}, \quad \frac{m_e v_e R_e}{\sqrt{1 - v_e^2}} = \frac{nh}{2\pi} \tag{1.3.A}$$

They describe the circular orbit of *the Bohr electron*,(6) of mass $m_e$ and negative charge $e_e$, round a source of positive charge, $e_n$. $h$ is Planck's constant and $v_e$ is the velocity of the electron at *the Bohr radius*, $R_e$,(6)

$$v_e = r\frac{\mathrm{d}\hat{\theta}}{\mathrm{d}t} \tag{1.3.B}$$

where we may assume the Bohr electron is in an inertial frame in, for example, $L_1$, and use special rather than general relativity.(6, 8) We call systems of this type *Bohr interactions*(6) if they are accompanied by an inverse distance-squared law for the force,

$$\frac{m_e}{\sqrt{1-v_e^2}}\frac{\mathrm{d}^2 r}{\mathrm{d}t^2}=\frac{e_e e_n}{r^2} \tag{1.3.C}$$

as in electromagnetism.[6, 25] The total energy, $E_e$, of the Bohr interaction[6] is

$$E_e = m_e\sqrt{1-v_e^2} \tag{1.3.D}$$

which we call *the internal energy*. Equations (1.3.A) through (1.3.D) hold in the frame of the source.

Provided $v_g \neq 1$, equations (1.2.J) and (1.2.K) allow,[4]

$$\frac{m_p\tilde{m}_s}{R}=\frac{m_p}{\sqrt{1-v_g^2}}v_g^2,\quad \frac{m_p}{\sqrt{1-v_g^2}}v_g R = n\frac{m_p\tilde{m}_s}{m} \tag{1.3.E}$$

where $m_p$ is the mass of the preon, and,

$$\tilde{m}_s=\frac{m_s}{\sqrt{1-v_g^2}} \tag{1.3.F}$$

Assuming that $m_p$ or $\tilde{m}_s$ or both are discrete, we may define a real positive constant we call *the versatile constant*,

[4] The first of equations (1.3.E) is identical to that derived from Newton's law of gravity [30] if we write it in the form,

$$\frac{m_p m_s}{R^2}=\frac{m_p v_g^2}{R}$$

The left-hand side is equal to the gravitational force and the right-hand side to the centrifugal force on the rotating body.

$$\frac{h_g}{2\pi} = \frac{m_p \tilde{m}_s}{m} \tag{1.3.G}$$

and equations (1.3.E) become

$$\frac{m_p \tilde{m}_s}{R} = \frac{m_p}{\sqrt{1-v_g^2}} v_g^2, \quad \frac{m_p}{\sqrt{1-v_g^2}} v_g R = \frac{nh}{2\pi} \tag{1.3.H}$$

while equations (1.2.J) and (1.3.F) imply,

$$\frac{m_p}{\sqrt{1-v_g^2}} \frac{\mathrm{d}^2 r}{\mathrm{d}t^2} = \frac{m_p \tilde{m}_s}{r^2} \tag{1.3.I}$$

Equations (1.3.H) and (1.3.I) are identical in form to equations (1.3.A) and (1.3.C). Interpreting $m_p \big/ \sqrt{1-v_g^2}$ as the kinetic energy of the preon, all are in the frame source. The internal energy is

$$E_I^{(l)} = m_p \sqrt{1-v_g^2} \tag{1.3.J}$$

from equation (1.3.D), where $l$ indexes the space, $L_l$, and $l=1$ here. We may represent our gravitational system in $L_1$ by a solution of *the versatile Dirac equation*,[(5-7)] with the potential given by Maxwell's equations for the photon.[(12)] By the Dirac equation, we shall always mean the versatile Dirac equation, which is identical to the original for half-angular behaviour of the spinor.[(5)] We use four-vector behaviour here.

The behaviour of the spinor phase is then the same as that of $\hat{\xi}$ in equations (1.2.O).[(6)] We suppose the two waves are identical, commenting on the wavelength in section 2.2. We calculate the exponential part of the wave function in $C_1$ using $L_1$ as a model,[(6)]

$$\Phi_C = \exp(i\xi) = \exp\left( im\frac{\tau}{r}\sqrt{\frac{m_s}{2r}} \right) \tag{1.3.K}$$

from equations (1.2.H) and (1.2.P).

We discuss the domain of equations (1.3.F), (1.3.G) and (1.3.H) if we rely on the Maxwell and Dirac equations providing a physical interpretation. Solving for $v_g$,

$$v_g = \frac{2\pi m_p \tilde{m}_s}{nh_g} \tag{1.3.L}$$

where $2\pi m_p \tilde{m}_s / h_g$ is the coupling constant. If $v_g > 1$ we may still solve the Maxwell and Dirac equations by a method that does not use perturbations, such as lattice gauge theory.[7, 8] $\tilde{m}_s$ and $m_p$ become imaginary if the other variables are real. A transition between a real and an imaginary $v_g$ corresponds to the transformation induced by using charge conjugation on the Dirac equation.[5, 7] $m_p$ and $m$ may be imaginary with the other variables real but $\tilde{m}_s$ and $m_s$ are negative. We also use *the tachyonic transformation*,[8] which gives physical meaning to the use of imaginary energies and momenta combined with the corresponding imaginary co-ordinates, or, equivalently, changes the signature of a space.[8] Here $v_g$ is inverted.

## 2. QUANTUM COSMOLOGY

### 2.1 The shell model

A circular symmetry became a spherical symmetry when the source was localised in the electromagnetic case.[6, 7] The Bohr electron was then confined to the surface of a sphere of radius $R_e$ centred on the source in *L*. The surface was formed from a superposition of the annular orbits of the Bohr electron, *the Bohr orbits*. Although a separate source is needed for an attractive interaction if it is electromagnetic, gravity has an attractive self-interaction, which we study by removing the source from the centre of the sphere. We call the result *the thin shell model*. We discuss both the internal and external appearance of such a shell, supposing that both obey equations (1.3.H) and (1.3.I).

From *the internal view*, the interior of the shell is a new space, $L_0$, with a metric,

$$\mathrm{d}\hat{\tau}_0^2 = \mathrm{d}t^2 - \mathrm{d}r^2 - r^2\left(\mathrm{d}\vartheta^2 + \sin^2\vartheta\,\mathrm{d}\varphi^2\right) \tag{2.1.A}$$

where the shell is addressed by the spherical polar co-ordinates $(r, \vartheta, \varphi)$ and the surface at $r = R$ consists of a superposition of annuli containing the Bohr orbits of the preon. The equivalent of $\tilde{m}_s$ and $m_p$ in $L_0$ are

$$\tilde{m}_s = \tilde{m}_s^{(0)}, \quad m_p = 2\tilde{m}_s^0 \tag{2.1.B}$$

where $2\tilde{m}_s^{(0)}$ is the rest mass of the shell and the superscript on $\tilde{m}_s$ will always indicate the space. The factor two is inserted because a small part, $2\delta\tilde{m}_s^{(0)}$, of the mass of the shell feels the attraction of the rest acting at the centre of mass and then appears twice in a summation, so that the total attractive force is proportional to $2\tilde{m}_s^{(0)2}$. The velocity of rotation of each annulus is

$$v_S = \sqrt{\frac{\tilde{m}_s^{(0)}\sqrt{1-v_S^2}}{R}} = \sqrt{\frac{\tilde{m}_s^{(0)}}{2R^2}\left(-\tilde{m}_s^{(0)} + \sqrt{\tilde{m}_s^{(0)2} + 4R^2}\right)} \tag{2.1.C}$$

from equations (1.3.H). We relate the internal view to the Bohr interaction we studied in section 1, *the external view*, by supposing that the internal energy is

$$E_I^{(0)} = m_s = 2\tilde{m}_s^{(0)}\sqrt{1-v_S^2} \tag{2.1.D}$$

where the last equality comes from equations (1.3.J) and (2.1.B). We separate $m_s$ into a kinetic and potential term,[6]

$$m_s = \frac{2\tilde{m}_s^{(0)}}{\sqrt{1-v_S^2}} - \frac{2\tilde{m}_s^{(0)2}}{R} \tag{2.1.E}$$

from equations (1.3.H) and (2.1.B). The exterior of the shell is a four-dimensional version of $C_1$, which we shall also call $C_1$, and we suppose the metric in equation (1.1.C) holds with additional terms denoting spherical symmetry,

$$\mathrm{d}\tau^2 = \frac{2m_s}{r}\mathrm{d}s^2 - \frac{r}{2m_s}\mathrm{d}r^2 - r^2\left(\mathrm{d}\vartheta^2 + \sin^2\vartheta\,\mathrm{d}\varphi^2\right) \tag{2.1.F}$$

The terms in the bracket have no effect on the conclusions in section 1 if $\mathrm{d}\vartheta$ and $\mathrm{d}\varphi$ are zero, although they do mean that the metric is not a solution of the Einstein equation in general.[25, 28] The preon, with any mass $m_p$ allowable, is now distinct from the original shell and,

$$\tilde{m}_s^{(1)} = 2\tilde{m}_s^{(0)}\sqrt{\frac{1-v_S^2}{1-v_g^2}} \tag{2.1.G}$$

from equations (1.3.F) and (2.1.D).

In section 1 we moved from a description using rotational motion in a spatial frame in $L_1$ to the language of curvature in $C_1$. The first uses a step function factor, $+\sqrt{1-\delta(r-R)v_g^2}$ or its inverse, to change the size of $\mathrm{d}t$ and $r\,\mathrm{d}\hat{\theta}$ in the moving frame on the annulus itself. The second uses a smooth variation of the factor $2m_s/r$ or its inverse to change the size of $\mathrm{d}s$ and $\mathrm{d}r$. The interchangeability is even clearer if we compare $L_0$ and $C_1$. We pick one of the annuli for the internal view and address it with polar co-ordinates $(r,\theta_0)$ in the alternative metric for $L_0$,

$$\mathrm{d}\hat{\tau}_0^2 = \mathrm{d}t^2 - r^2\,\mathrm{d}\theta_0^2 - \mathrm{d}r^2 - \mathrm{d}\hat{x}_3^2 \tag{2.1.H}$$

where $r = R$ on the annulus. Then from equations (1.3.H), (2.1.C) and (2.1.G),

$$v_g = v_S\sqrt{2} \tag{2.1.I}$$

and so we may set,

$$\theta_0 = \theta \tag{2.1.J}$$

from integration using equations (1.1.G) and (1.2.I) and the definition of $v_S$. In this sense the internal and external views describe the same interaction and we could hope to describe the transition from one to the other using co-ordinate transformations respected by the Maxwell and Dirac equations. We may pass from the internal view in the frame of the source to the dashed frame of the preon using a Lorentz transformation in $L_0$,

$$\mathrm{d}t, \mathrm{d}\theta_0, \mathrm{d}\hat{r} \rightarrow \mathrm{d}t', \mathrm{d}\theta_0', \mathrm{d}\hat{r}' \tag{2.1.K}$$

Since we have applied equations (1.3.H) and (1.3.I) to a spatial orbit for the internal view but a temporal orbit for the external view we add a tachyonic transformation to the double-dashed frame in $L_0$,

$$\mathrm{d}t', \mathrm{d}\theta_0', \mathrm{d}\hat{r}' \rightarrow \mathrm{d}t'', \mathrm{d}\theta_0'', \mathrm{d}\hat{r}'' \tag{2.1.L}$$

We assess how far this is born out. We transform the internal energy, $m_s$, to the frame of the preon,[6, 8]

$$m_s \rightarrow \left(2\tilde{m}_s^{(0)}, E_X^{(0)}, 0, 0\right), \quad E_X^{(0)} = 2\tilde{m}_s^{(0)} v_S \tag{2.1.M}$$

from equation (2.1.D). $E_X^{(0)}$ becomes *the external energy* if we add the tachyonic transformation,

$$\left(2\tilde{m}_s^{(0)}, E_X^{(0)}, 0, 0\right) \rightarrow \left(E_X^{(0)}, 2\tilde{m}_s^{(0)}, 0, 0\right) \tag{2.1.N}$$

where we have assumed a changed signature rather than using imaginary co-ordinates explicitly. We extend this to $L_1$,

$$E_X^{(l)} = m_p v_g \tag{2.1.O}$$

where $l = 1$. If $s$ is ignorable,

$$2\gamma = \frac{\partial L}{\partial \tilde{s}}, \quad \tilde{s} = \frac{\mathrm{d}s}{\mathrm{d}\tau} \tag{2.1.P}$$

where $\gamma$ is a constant of the motion,[28] $\tau$ is the interval and $L$ is the Lagrangian. For the metrics in equations (2.1.C) and (2.1.F),

$$\gamma = \frac{E_X^{(1)}\sqrt{2}}{m_p} \tag{2.1.Q}$$

if $\mathrm{d}\,r = \mathrm{d}\vartheta = \mathrm{d}\varphi = 0,$ from equations (1.1.J), (1.3.F), (1.3.H) and (2.1.O). We notice that the constant momentum, $2\tilde{m}_s^{(0)}$, in relation (2.1.N) is not represented by general relativity and we too will consider it no further.

The gravitational interaction of thin shells has been studied extensively,[1, 9-11, 16, 21-23, 26] where we have chosen some citations whose treatment is particularly appropriate. A motion along a great circle at every point on a classical shell is impossible and the literature discusses radial motion. The quantum treatment we proposed above is new in that we suppose the motion is tangential to a circular orbit but a connection with atomic spectra has already been noticed,[11] after using the standard method[18] of quantisation, which we call *s-quantisation*. Kuchar[26] finds that the interior of the shell is flat[5] in the classical case and that the total energy of the shell is

$$m_s' = \frac{e_S^2}{2R'} - \frac{2\tilde{m}_s^{(0)2}}{R'} + \frac{2\tilde{m}_s^{(0)}}{\sqrt{1-(v_S')^2}} \tag{2.1.R}$$

where $e_S$ is the electric charge on the shell, $R'$ is the radius of the shell surface and,

[5] See Kuchar[26] and also Frolov and Novikov[20] for qualifications on signatures.

$$v_S' = \frac{\mathrm{d}R'}{\mathrm{d}t} \tag{2.1.S}$$

is the velocity of expansion or contraction of the shell. Setting $e_S = 0$, $v_S' = v_S$ and $R' = R$, we find $m_s' = m_s$ from equation (2.1.E), which implies the existence of a force that obeys equation (1.3.I) applied to $L_0$.[6] Kuchar quotes the usual Schwarzschild solution[19, 25, 28] to the Einstein equation,

$$\mathrm{d}\tau_2^2 = \left(1 - 2m_s'/r\right)\mathrm{d}s_2^2 - \frac{\mathrm{d}r^2}{\left(1 - 2m_s'/r\right)} - r^2\left(\mathrm{d}\vartheta^2 + \sin^2\vartheta\,\mathrm{d}\varphi^2\right) \tag{2.1.T}$$

where $s_2$ is the temporal co-ordinate, as the metric for the exterior of the shell. This is only the same as the metric in equation (2.1.F) in the limit $r \to 0$ with $m_s$ negative.[20]

## 2.2 The shell universe

We model the universe as a thin shell, called *the unishell*. Applying equation (2.1.F), the metric for the unishell in $C_1$ is

$$\mathrm{d}\tau_u^2 = \frac{2m_u}{r_u}\mathrm{d}s_u^2 - \frac{r_u}{2m_u}\mathrm{d}r_u^2 - r_u^2\left(\mathrm{d}\vartheta^2 + \sin^2\vartheta\,\mathrm{d}\varphi^2\right) \tag{2.2.A}$$

where $r_u$ is the radial and $s_u$ the temporal co-ordinate, $m_u$ the total energy of the unishell and,

$$\frac{2m_u}{R_u} = 1 \tag{2.2.B}$$

with $r_u = R_u$ at the Bohr radius, the position of the observer. The orbital velocity of the shell is

$$v_u = \frac{1}{\sqrt{2}} \tag{2.2.C}$$

from equations (1.2.J). We suppose the gravitational charge, $m_u$, generates the space and supplies the vacuum energy, $E_X^{(1)} = m_p v_u$ in this context, from equation from equation (2.1.O). $m_u$ and $R_u$ are arbitrary, apart from the condition in equation (2.2.B), and $m_u$ is not to be confused with the mass of the universe calculated using other models from the classical theory. We add a second thin shell, *the copyshell*, which generates the metric in equation (2.1.F) in $C_1$ in the absence of other shells. We suppose that $v_u$ and $v_g$ may remain unchanged by the addition of the copyshell or that they may change, each alternative giving rise to a different model. We represent both possibilities as hierarchies, starting with the first.

Our discussion already constrains the relation between the internal view of a shell in $L_0$ and the external view in $C_1$ and also shows that we may map the external view to a flat space, $L_1$. We suppose this is the internal view of another shell. The new shell has an external view in a curved space $C_2$, which we may map to a flat space $L_2$, if we apply v-quantisation in $L_2$ and not $L_1$. The relation between the internal and external view of the new shell is the same as it was for the original shell. We may repeat this procedure, downwards as well as upwards, obtaining the sequences in table 1,

**Table 1.**

**The original hierarchies for the copyshell and unishell**

| **Space** $L_l, C_l$ | **Source Mass** | **Name** | **Internal Energy** | **Speed in** $L_l$ | $n$ & $m$ |
|---|---|---|---|---|---|
| $L_{-1}, C_{-1}$ | $m_s/4$<br>$m_u/4$ | $\begin{Bmatrix}\text{protoshell, external}\\ \text{multishell, internal}\end{Bmatrix}$<br>? | $\sqrt{1-v_S^2/2}$<br>$\sqrt{7/8}$ | $v_S/\sqrt{2}$<br>$1/2\sqrt{2}$ | $n$,<br>$m/2$<br>? |
| $L_0, C_0$ | $m_s/2$<br>$m_u/2$ | $\begin{Bmatrix}\text{multishell, external}\\ \text{copyshell, internal}\end{Bmatrix}$<br>unishell, internal<br>? | $\sqrt{1-v_S^2}$<br>$\sqrt{3/4}$ | $v_S$<br>$1/2$ | $n$,<br>$m/\sqrt{2}$<br>? |
| $L_1, C_1$ | $m_s$<br>$m_u$ | $\begin{Bmatrix}\text{copyshell, external}\\ \text{combishell, internal}\end{Bmatrix}$<br>unishell, external<br>? | $\sqrt{1-v_g^2}$<br>$1/\sqrt{2}$ | $v_g =$<br>$v_S\sqrt{2}$<br>$1/\sqrt{2}$ | $n$<br>$m$<br>? |
| $L_2, C_2$ | $2m_s$ | $\{\text{combishell, external}\}$ | $\sqrt{1-2v_g^2}$ | $v_g\sqrt{2}$ | $n$,<br>$m\sqrt{2}$ |

The names in brackets show the sequence of shells arising from the copyshell, with external or internal marking the view. The ? in the names columns leave the nature of the other levels of the unishell open. The initial value in the second, fourth and fifth columns applies to the copyshell and the last to the unishell. The values in the second column give the total energy of the source and those in the fourth column multiplied by the mass of the preon give the internal energy. The first two values in the sixth column give the quantum numbers *n* and *m* for the copyshell hierarchy, where we have applied v-quantisation to $L_1$, and the ? signify that only the ratio is known for the unishell. *R* is constant throughout. The second column comes from equations (1.2.J) applied to the fifth column, the fourth column comes from equations (1.3.J), the fifth from equation (2.1.I) and the sixth from equation (1.2.K), where we do not insist on *m* being integer. All are applied in $L_l$. We cannot apply equations (1.3.H) to the unishell hierarchy in $L_2$ and we have not included it.

The hierarchies in table 1 are scaling if we represent the mass of the source as an area. We may plot them as tiles in a tiling hierarchy, an example of *a tesseral hierarchy*.[3, 17] One with this branching ratio was first explored by Lusby Taylor.[27] At each level in table 1 the preon is accompanied by the wave described in sections 1 and 2. We may see the hierarchies as generated by the wave and we describe the method for the copyshell. After a time $t = T$, when the preon has completed one circuit of the orbit in $L_0$, the phase of the wave has completed *n* circuits in $L_0$, from equation (1.2.O) and *m* circuits in $C_1$, from equations (1.2.P) and (2.1.J). In section 2.1 we used general relativity to combine $L_0$ and $C_1$ into a single space we will call $E_{l,l+1}$, where *l* indexes the spaces $L_l$ and $C_{l+1}$ and $l = 0$ here. Since the orbit in $L_0$ is spatial, while the orbit in $C_1$ is temporal, the velocity of the wave in $E_{0,1}$, is[6]

$$v_g^w = \frac{1}{v_g} \tag{2.2.D}$$

from equation (1.2.K). Transforming the mass of the source, $(m_s)_1$, and preon, $(m_p)_1$, and the Bohr radius, $(R)_1$, in $L_1$ and $C_1$,

$$(m_s)_1 = mm_s, \quad (m_p)_1 = \frac{m_p}{m}, \quad (R)_1 = mR \tag{2.2.E}$$

while leaving equations (1.3.F), (1.3.G) and (1.3.H) and $\mathrm{d}\tau/\mathrm{d}t$ unchanged, the wave in $C_1$ will take the same time, $t = T$, to complete one circuit of the orbit as the preon in $L_0$. Instead of identifying this preon with the preon in $C_1$ as we did in section 2.1, we identify the preon in $L_l$ with the wave in $C_{l+1}$. This automatically includes the exchange of a spatial and the temporal co-ordinate in the map (2.1.L).[7] We would like to suppose, in addition to using our original argument in section 1 to relate the preon in $L_l$ and $C_l$, that the preon in $L_l$ and $L_{l'}$, $l \neq l'$ are the same. If we use arguments which only rely on the relationships between levels and not on their absolute value, we may do this for a tesseral hierarchy, since it is scaling. Iteration of the relation in equation (2.2.E) shows that we may add,

6 Louis de Broglie also associated a wave, the pilot wave, with a particle and derived this relationship between their velocities.[31]

7 The wave in $L_0$ now has the same wavelength as the wave first predicted by Louis de Broglie.[6, 31]

$$(m_s)_l = 2m\left(\sqrt{2}\right)^{l-1}(m_s)_{l-1},\quad (m_p)_l = \frac{(m_p)_{l-1}}{m\left(\sqrt{2}\right)^{l-1}},\qquad (2.2.F)$$

$$(R)_l = m\left(\sqrt{2}\right)^{l-1}(R)_{l-1}$$

These describe series, $l = 2,3,4\ldots$ with the term for $(\cdot)_l$ given by the term for $(\cdot)_{l-1}$. The first and third series describe a tesseral hierarchy with a branching ratio of $2m$ and $m$, respectively, interleaved with one whose branching ratio is the same as the square root of the branching ratio of the Lusby Taylor hierarchy. We may also reverse the direction of a tesseral hierarchy(4) so that the second series is the same as the third. We shall restrict our series to the one made explicit in table 2 for all the algebraic calculations that follow.

We continue with the first model. The presence of two shells, the unishell and copyshell, destroys the spatial symmetry and we suppose that the wave function of both collapse to a single annulus in orthogonal subspaces of $L_1$. The unishell then combines with the copyshell to form the combishell. A preon, on the shell surface, as it always will be from now on, acquires the rotational motion of both the original orbits round the unishell and copyshell, with the first a contribution to the energy and the second to the momentum. The energy-momentum four-vector is $\left(m_p v_u, m_p\sqrt{m_s/R},0,0\right)$ from equations (1.2.J) and (2.1.O), where we have chosen a sign. Transforming to the rest frame, the total energy is

$$m_u' = m_p\sqrt{(v_u)^2 - \frac{m_s}{R}},\qquad (2.2.G)$$

Adopting co-ordinates showing spatial symmetry and setting $m_u = R/2$ and $r = R,$ the metric in $C_2$ becomes

$$\mathrm{d}\tau_2^2 = \left(1 - 2m_s/r\right)\mathrm{d}s_u^2 - \frac{\mathrm{d}r^2}{1 - 2m_s/r} - r^2\left(\mathrm{d}\vartheta^2 + \sin^2\vartheta\,\mathrm{d}\varphi^2\right) \quad (2.2.\mathrm{H})$$

from equations (2.1.P), (2.1.Q), (2.2.B) and (2.2.C) with $E_X^{(1)} = m_u'$ and $\mathrm{d}r = \mathrm{d}\vartheta = \mathrm{d}\varphi = 0.$ This agrees with the metric in equation (2.1.T) if $s_u = s_2.$ Equation (1.3.I), continues to hold,[(28)] which means that the interaction can be represented by the Dirac equation, omitting the quantisation implied by the second of equations (1.3.H).[(6, 8)] We call this *Planck quantisation* or *p-quantisation*. It is weaker than v-quantisation, implying only that *n* is integer. The Dirac equation becomes classical in the absence of p-quantisation.[(8)]

For the second model, we suppose the two shells form a single hierarchy at every scale. This is our final rendition of Mach's principle.[(19)]

**Table 2.**

**The hierarchy showing the many guises of a single shell**

| **Space** $L_l, C_l$ | **Source & preon** | **Name** | **Speed** | $n$ & $m$ | $h_g/2\pi$ |
|---|---|---|---|---|---|
| $L_{-1}, C_{-1}$ | $m_s/4$<br>$\frac{m_s/2}{\sqrt{1-v_S^2/2}}$ | protoshell, external<br>multishell, internal | $v_S/\sqrt{2}$ | $n$<br>$\frac{m}{2}$ | $\frac{m_s^2}{4m(1-v_S^2/2)}$ |
| $L_0, C_0$ | $m_s/2$<br>$\frac{m_s}{\sqrt{1-v_S^2}}$ | multishell, external<br>copyshell, internal | $v_S$ | $n$<br>$\frac{m}{\sqrt{2}}$ | $\frac{m_s^2}{m\sqrt{2}(1-v_S^2)}$ |
| $L_1, C_1$ | $m_s$<br>$\frac{2m_s}{\sqrt{1-v_g^2}}$ | copyshell, external<br>unishell, internal | $v_g =$ $v_S\sqrt{2}$ | $n$<br>$m$ | $\frac{2m_s^2}{m(1-v_g^2)}$ |
| $L_2, C_2$ | $2m_s$<br>$\frac{4m_s}{\sqrt{1-2v_g^2}}$ | unishell, external<br>combishell, internal<br>(observer) | $v_g\sqrt{2}$ | $n$<br>$m\sqrt{2}$ | $\frac{4m_s^2\sqrt{2}}{m(1-2v_g^2)}$ |
| $L_3, C_3$ | $4m_s$<br>$\frac{8m_s}{\sqrt{1-4v_g^2}}$ | combishell, external | $2v_g$ | $n$<br>$2m$ | $\frac{16m_s^2}{m(1-4v_g^2)}$ |

In addition to the total energy of the source, the second column contains the mass of the preon. The entries are chosen to satisfy equations (2.1.B) and (2.1.D). $h_g$ satisfies equations (1.3.F) and (1.3.G) and varies between rows.

Instead of using the gravitational constant of the motion $\gamma$, we may deduce the metric by supposing that curvature is caused by a rotational motion, directly, and using the Lorentz transformations. We introduce a space $\underline{L}$ with a metric,

$$\mathrm{d}\underline{\tau}^2 = \mathrm{d}\underline{s}^2 - \mathrm{d}\underline{r}^2 \tag{2.2.I}$$

and a stationary frame, $F_S$, with co-ordinates $(\underline{s}, \underline{r})$. A frame $F_M$ with co-ordinates $(\underline{\underline{s}}, \underline{\underline{r}},)$ is moving with velocity $v$ with respect to $F_S$. A time interval in the moving frame, $\mathrm{d}\underline{\underline{s}}$, suffers a dilation while a spatial interval parallel to the direction of motion, $\mathrm{d}\underline{\underline{r}}$, suffers a Lorentz contraction,[19]

$$\mathrm{d}\underline{\underline{s}} = \frac{\mathrm{d}\underline{s}}{\sqrt{1-v^2}}, \quad \mathrm{d}\underline{\underline{r}} = \mathrm{d}\underline{r}\sqrt{1-v^2} \tag{2.2.J}$$

Substituting in equation (2.2.I),

$$\mathrm{d}\underline{\tau}^2 = \left(1-v^2\right)\mathrm{d}\underline{\underline{s}}^2 - \frac{\mathrm{d}\underline{\underline{r}}^2}{\left(1-v^2\right)} \tag{2.2.K}$$

Specialising in the space $C_2$ by setting $v = v_g\sqrt{2}$, $\underline{\underline{s}} = s_u$ and $\underline{\underline{r}} = r$, using equations (1.3.F) and (1.3.H) and adopting co-ordinates showing spatial symmetry, we recognise the metric in equation (2.2.H). To retrieve the Schwarzschild metric, we have made $\mathrm{d}\underline{\underline{r}}$ an interval along the radius but the motion we have studied so far is tangent to a circular orbit. We conjecture that *a Lorentz contraction takes place along the radius for a quantum rotational motion.* We give our reasons. Since general relativity predicts it, we know that the conjecture holds if the rotational motion is caused by gravitational attraction. We may put the equations for electromagnetism, (1.3.A), and (1.3.C), into the same form as the equivalent for gravity, (1.3.H) and (1.3.I), by setting the charge on the source, $e_n'$, and Bohr electron, $e_e'$, to,

$$e_n' = \frac{e_e e_n}{m_e}, \quad e_e' = m_e \tag{2.2.L}$$

We may then argue that the Maxwell and Dirac equations hold with the dashed values in the same way as we did previously.[7, 8] We may also offer the same explanatory framework as we have done for general relativity. The conjecture then supposes that we associate the curvature with the motion in the quantum case, rather than the field. However, we have not explored generalising this formulation of QED-theory to other gauge fields.

## 3. QED THEORY

### 3.1 The full solution for gravity

The inverse-distance relation between a source and the potential leads to Maxwell's equations, as we found. [7] The equations, in the form defining the gravitational four-vector potential, $A_\mu$, become Poisson's equation and are still exact if $A_\mu$ and the gravitational charge density four-vector, $\rho_\mu$, are scalars, $A_0$ and $\rho_0$, that do not depend on time.[12, 30] Poisson's equation is the differential form of Newton's law of gravity if, in addition, we allow $A_0$ and $\rho_0$ to vary with time.[30] Dirac's equation describes the movement of a particle in the potential, $A_\mu$, and has a classical form, which could be described as a relativistic representation of circular motion.[8] The Maxwell and Dirac equations are obeyed for the interaction between a particle and an arbitrary distribution of source density, $\rho_\mu(x_\mu)$, if Bohr's equations are obeyed,[7] as they are in $L_1$ if we leave out the coupling to the unishell. We need to transform to the tachyon frame to compare the result with our experience of time but then proceed as we discuss in more detail for $L_2$, where Bohr's equations are also obeyed, if we include the coupling to the unishell.

We move from $L_1$ to $L_2$, using table 2,

$$\overline{m}_s' = 2m_s, \quad \overline{m}_p' = \frac{4m_s}{\sqrt{1-2v_g^2}}, \qquad (3.1.A)$$
$$\overline{v}_g = v_g\sqrt{2}, \quad \frac{\overline{h}_g}{2\pi} = \frac{4m_s^2\sqrt{2}}{m\left(1-2v_g^2\right)}$$

where $\overline{m}_s'$ is the mass of the source, $\overline{m}_p'$ and $\overline{v}_g$ the mass and velocity of the preon and $\overline{h}_g/2\pi$ the versatile constant. We transform $\overline{m}_p'$ to $\overline{m}_p$, which is the same for every source and velocity,

$$\overline{m}_s = \frac{2m_s^2}{\sqrt{1-2v_g^2}}, \quad \overline{\tilde{m}}_s^{(2)} = \frac{2m_s^2}{1-2v_g^2}, \quad \overline{m}_p = 4, \quad \overline{R} = \frac{m_s R}{\sqrt{1-2v_g^2}} \qquad (3.1.B)$$

where we have transformed $\overline{m}_s'$ to $\overline{m}_s$, $\tilde{m}_s^{(2)}$ to $\overline{\tilde{m}}_s^{(2)}$ and $R$ to $\overline{R}$ to satisfy equations (1.3.F), (1.3.G) and (1.3.H). Our new variables, $\overline{\tilde{m}}_s^{(2)}$, $\overline{m}_p$, $\overline{v}_g$, $\overline{h}_g/2\pi$ and $\overline{R}$ are now in the form required to find the dynamics of the preon in $L_2$ for an arbitrary distribution, $\bar{\rho}_\mu(x_\mu)$, of source density.[7] We reserve $\rho_\mu$ for a source and Bohr radius of $m_s$ and $R$, respectively, while the equivalent for $\bar{\rho}_\mu$ are $\bar{\bar{m}}_s^{(2)}$ and $\bar{R}$. Then we put the Maxwell and Dirac equations on a lattice and show that the result is scaling,[7, 8] in the sense normally understood for renormalisation.[29] For this and for renormalisation itself, we suppose that the bare gravitational and inertial masses are independent, inserting the physical equality at the end. We can use either the computational methods of lattice gauge theory[7, 8] for renormalisation or, for small velocities, the propagator approach. The Dirac equation with a source term[29] is not explicitly covariant for four-vector behaviour of the mass term, although we may choose any one frame and insert a scalar value equivalent to the rest mass.[5] We find the propagator in this frame and then the wave function, which does satisfy the Dirac equation with joint four-vector behaviour of the wave function and mass term.

In general relativity, the mass density, $\rho_\mu$, becomes the second rank energy-momentum-stress or *matter tensor*, $T_{\mu\nu}$.[28] If $T_{\mu\nu}$ is in the form of the direct product, $v_\mu v_\nu$, of two equal four-vectors, the Einstein tensor, $G_{\mu\nu}$,[28] is necessarily of this form too and we may decompose the Einstein equation into a four-vector equation. We define a map,

$$T_{\mu\nu} \leftrightarrow \rho_\mu \tag{3.1.C}$$

and provide the map explicitly in the rest frame,

$$T_{00} = \rho_0 \tag{3.1.D}$$

Equations (3.1.B) can then be used to provide a map, $T_{\mu\nu} \leftrightarrow \overline{\rho}_\mu$. We call matter tensors that can be mapped in this way *simple matter tensors* and the rest *composite matter tensors*. Since $G_{\mu\nu}$ may be found from the metric tensor, $g_{\mu\nu}$,[25, 28] it will be sufficient to discuss the latter. We may obtain $g_{\mu\nu}$ from the energy-momentum four-vector of the Bohr interaction, $E_\mu(x_\mu)$, which becomes the internal energy, $E_I^{(2)}$, in our rest frame. $E_\mu$ is found by using the appropriate operator on the wave function.[6, 7] We define a map,

$$g_{\mu\nu} \leftrightarrow E_\mu E_\nu \tag{3.1.E}$$

with an explicit map in the rest frame,

$$g_{00} = \frac{E_0^2}{\overline{m}_p^2}, \quad g_{11} = -\frac{\overline{m}_p^2}{E_0^2}, \quad g_{22} = -r^2, \quad g_{33} = -r^2 \sin^2\vartheta \tag{3.1.F}$$

with the other elements zero, from the metric in equation (2.2.H) and equations (1.3.F), (1.3.H), (1.3.J) and (3.1.A) applied in $L_2$.

We have shown that QED theory can be used to provide a renormalisable quantum version of general relativity for simple matter tensors. We discuss the implications of our results for composite matter tensors by proposing a thought experiment, which, with modifications, could also be performed in practice using a computer simulation. We may calculate a weighted sum of the simple matter tensors at a single point, $x_i$, in $L_1$ for a given statistical distribution of sources to generate the composite matter tensor at that point, $T_{\mu\nu}(x_i)$. We may neglect the effects of curvature on the properties of the statistical distribution.(19, 28) We enlarge. The statistical distribution of simple matter tensors required at $x_i$ is

$$T_{\mu\nu}(x_i) = \sum_{k=0}^{j} f_k(x_i) v_{k\mu}(x_i) v_{k\nu}(x_i) \qquad (3.1.G)$$

where the $f_k$ are the weights and $j$ may be any positive integer. We require that the equations have a solution for some $v_{k\mu}$. Martin(28) is instructive on some simple solutions when $T_{\mu\nu}$ can be diagonalised. Suitable sets, $\{T_{\mu\nu}(x_i)\}$, may then be input to the Einstein equation to find the curvature and from this the dynamics of the preon.(19, 28) We may simulate the same statistical distribution, $T_{\mu\nu}(x_i)$, at $x_i$ with the sources, $v_\mu(x_{ik})\sqrt{f(x_{ik})}$, at points, $x_{ik}$, which we place near $x_i$ in $L_1$. These produce a set of simple matter tensors, $\{f(x_{ik})v_\mu(x_{ik})v_\nu(x_{ik})\}$, as we vary *k* or *i*. We transform the $\{v_\mu(x_{ik})\sqrt{f(x_{ik})}\}$ to $L_2$ and the form required by the Maxwell and Dirac equations and use these to find the dynamics of the preon. The dynamics found, taken to the classical limit and suitably averaged, should agree with those found by the previous method if the statistical distributions for the $x_i$ are the same. This means that the behaviour of the Einstein equation, for the composite matter tensors that can be derived in this way from simple ones, depends only on the latter.[8]

[8] Clearly, we have also used classical statistical thermodynamics.

### 3.2 Bohr's equations and quantisation

We have now given several accounts, here and previously,[5-8] of systems that obey Bohr's equations. We draw the threads together. We started with electromagnetism and described a system with a spatial circular orbit that obeyed Bohr's equations.[6] This took place in a flat space, one of *S*, *M* or *L*.[6-8] We showed that Bohr's equations were still obeyed if we exchanged and complexified the temporal co-ordinate and a spatial co-ordinate defining the orbit.[8] We have shown here that Bohr's equations are also obeyed if we transfer to a curved space, $C_1$. The addition of curvature permits us to describe the system using only the co-ordinates defining the orbit. Since we know that the quantum theory is QED if Bohr's equations are obeyed,[7, 8] QED theory applies to all these systems.

It may be wondered what has happened to the transition from a classical to a quantum theory, since the standard method replaces variables with operators obeying commutation or anticommutation relations.[18] We call this s-quantisation. The underlying equations on which our method depends are Bohr's. Bohr's first equation is an approximation of Maxwell's theory,[12] joining an inverse distance-squared law for an electrical force to the Newtonian equation for a relativistic circular motion.[14] Bohr's second equation for Planck quantisation was considered new.[13] Both together seem at first sight to form a new theory, the old quantum theory, as it emerged in Bohr's hands.[13-15] Appearances are deceptive. The quantum condition we use for Planck quantisation in section 2.2 originally arose from the requirement that the spatial part of a wave function be single-valued for a bound state.[(6)] However, we have now shown that only two-dimensional co-ordinates, for example, $\tau$ and $r$ or $r\,\mathrm{d}\theta$ and $r$, are required and so we insist on no more than a single-valued wave function.

The reflector wave function can be seen as a new generalisation of the quaternion algebra[(2)] in which a Lorentz transformation, implemented by a temporal rotation, mirrors a spatial rotation,[(5)]

$$\begin{aligned} &\mathbf{Q}' = \mathbf{R}_S \mathbf{Q} \mathbf{R}_S^{\ddagger}, \qquad && \underline{\mathbf{Q}'} = \mathbf{R}_S| \underline{\mathbf{Q}}\, \mathbf{R}_S|^{\ddagger}, \\ & && \underline{\mathbf{Q}'} = \mathbf{R}_T| \underline{\mathbf{Q}}\, \mathbf{R}_T|^{\ddagger}, \\ &\mathbf{R}_S| = \mathbf{R}_S|\left(\mathbf{R}_S, \mathbf{R}_S\right), && \\ &\mathbf{R}_T| = \mathbf{R}_T|\left(\mathbf{R}_T, \mathbf{R}_T^{\ddagger}\right) && \end{aligned} \tag{3.2.A}$$

where all the variables are functions of quaternions and $\mathbf{R}_T$ has a unit modulus. The first and second equations describe the same spatial rotation while the third describes a temporal rotation. With this algebra the spin of a fermion varies like a four-vector. This means that the wave function can describe the positions of both the original particle and a second particle in addition and leads to the hierarchical system of particles in table 2.[9]

We have arrived at an economical redefinition of Bohr's second equation; that the wave must have a single value at each point. The usual distinction between a quantum and a classical system has gone and, yet, it is still possible to derive QED theory in its usual form.[(7, 8)] We examined general relativity to see if this classical theory was sufficiently like the classical limit of quantum electrodynamics for us to apply Bohr's equations to the former as well as the latter. We found that an hierarchy of scaling versions of both theories was required, all compatible with Bohr's equations. Judicious choice of scale provides the same link, in like cases, between the Dirac current and the energy of the quantum system as that found in general relativity between the matter tensor and the metric. We have in this way applied QED-theory to general relativity without s-quantisation as the preliminary step.

[9] The usual argument for a wave, an appeal to interference, can be met by inserting Feynman's sum over histories applied to a particle whose unobserved course is indeterminable.

### 3.3 M-theory

String theory and its derivative M-theory is based on the assumption that the building block of the physical world is the string.[24] Like string theory, instead of discussing two concepts, a field or curvature, which then acts on some unrelated particle, we discuss the interaction itself. We show that M-theory and QED theory could converge with this common approach. We have substituted excitations of a topological loop for a bound fermion.[6, 8] We may also include the winding number and arrive at the complete expression for the energy of a compactified string, $E$.[24] Using $l = 1$ as an example, we define,

$$E = E_X^{(1)} + m_s \tag{3.3.A}$$

where $E_X^{(1)}$ and $m_s$ are the external and internal energies of two different hierarchies, which we shall call the twinshell and the copyshell, respectively. Using QED-theory, we know that,[7, 8]

$$E_X^{(1)} = m_p \frac{n_r}{n_\theta}, \quad m_s = m_p \sqrt{1 - v_g^2} \tag{3.3.B}$$

where $m_p$ is the mass of the Bohr electron and preon and we use equation (1.3.J) and our previous result and terminology[8] for the excitation numbers, $n_r$ and $n_\theta$. The second equation provides the principle energy levels and the first adds the fine structure in the familiar electromagnetic spectrum of the hydrogen atom.[8, 18] From equation (2.1.O),

$$E_X^{(1)} = m_p v_t \tag{3.3.C}$$

where $v_t$ is the velocity of the twinshell. From equations (1.3.H),

$$m_p v_t \tilde{R}_t = n_r \frac{h_g}{2\pi}, \quad \tilde{R}_t = \frac{R_t}{\sqrt{1 - v_t^2}} \tag{3.3.D}$$

where $\tilde{R}_t$ is the Bohr radius of the twinshell in the rest frame of the preon. Using equation (3.3.C), the external energy becomes

$$E_X^{(1)} = \frac{n_r h_g}{2\pi \tilde{R}_t} \tag{3.3.E}$$

From equations (1.3.H) for the copyshell,

$$m_p v_g \tilde{R} = n_\theta \frac{h_g}{2\pi}, \quad \tilde{R} = \frac{R}{\sqrt{1 - v_g^2}} \tag{3.3.F}$$

where $\tilde{R}$ is the Bohr radius in the rest frame of the preon. From equations (3.3.B) and (3.3.E),

$$\tilde{R} = \frac{\tilde{R}_t}{v_g} \tag{3.3.G}$$

We cannot represent these two shells as a single hierarchy of the type shown in table 2 because instead of *n* there are two parameters, $n_\theta$ and $n_r$. We may still build a single hierarchy with one parameter per level. We transform the mass of the preon and source to new values $m'_p$ and $m'_s$, respectively,

$$m'_p = m_p v_g, \quad m'_s = m_s \sqrt{1 - v_g^2} \tag{3.3.H}$$

where we have moved to the frame of the preon. We obtain,

$$m_s = \frac{m' h_g}{m_p \bar{v}_g \pi \sqrt{2}} \tag{3.3.I}$$

from equations (1.3.F), (1.3.G) and (3.1.A) and $m'$ is the new winding number for $m'_p$ and $m'_s$. The denominator on the right-hand side is $E_X^{(2)}$ for the copyshell hierarchy and the string energy, *E*, is now given with reference to two different levels of the original hierarchies. We may regard these as two levels of a new hierarchy, the first for $L_1$ with $n = n_r$ and entries for the twinshell and the second for $L_2$ with $n = n_\theta$ and entries for the copyshell. We treat each level in the original hierarchies similarly and interleave them in order, obtaining a new scaling hierarchy explicitly. We may also join the two originals to a third hierarchy, again different, if we iterate by the replacement,

$$m_p \bar{v}_g \rightarrow m_p \bar{v}_g + \overline{m}_s' \qquad (3.3.J)$$

in equation (3.3.I), where $\overline{m}_s'$ is the internal energy of the third hierarchy for $L_2$. $\overline{m}_s'$ introduces a new integer parameter and our procedure with equation (3.3.A) is repeated using equation (3.3.J). We return to finding the energy of the original string in equation (3.3.A). From equations (3.1.A), (3.3.F) and (3.3.I),

$$m_s = \frac{m'\tilde{R}}{n_\theta} \qquad (3.3.K)$$

From equations (1.2.K), (3.3.E) and (3.3.G), equation (3.3.A) becomes

$$E = \frac{n_r h_g}{2\pi \tilde{R}_t} + m''\tilde{R}_t, \quad m'' = \frac{m'}{m} \qquad (3.3.L)$$

where we insist that $m''$, the winding number for the string, is an integer, rather than $n_\theta$. Here we have replaced an unknown product, $\tilde{m}_s m_p$, with a ratio depending only on a velocity, from equations (1.3.F), (1.3.G) and (3.3.H).

Since we have not chosen the usual route to quantising a string,[24] our findings do not apply to any single quantised string theory but to their classical point of departure. However, M-theory not only rests on the same origin but is also held to signal the physical content in a common theory. We might therefore hope that our results comply with M-theory.

## 4. DISCUSSION

We have shown that in addition to half-angular behaviour of the wave function under rotation and its equivalent under Lorentz transformation, QED theory may be reproduced with four-vector behaviour.[(5)] Using this theory we recovered the usual result for a bound state of two point particles in which the potential followed an inverse-distance law.[(6, 8)] We generalised this[(7, 8)] to show that QED theory could be reduced to two conditions; firstly, that force varying as the inverse distance-squared should result in the circular motion of a particle, which, secondly, can be described by a field that is single-valued over a space.

We found that these conditions were compatible with general relativity. Both theories are relativistic and this leads to part of the latter theory having a quantum limit that can be described by the former. We have not shown that QED theory is the only possible quantum theory of gravity nor that a more complete quantum theory is impossible. Nor have we shown the relation between electromagnetism and gravity. We have only shown that they function in the same manner. The curvature usually associated with general relativity did not enter this result directly.

We have also shown that QED theory is likely to be related to M-theory as both rely on closed loops. It is hoped that a renormalisable quantum theory of gravity allied to M-theory might be of interest.

**ACKNOWLEDGEMENTS**

One of us (Bell) would like to acknowledge the assistance of E.A.E. Bell. She would like to dedicate this paper to Thomas Bertram Radley, who originally conceived the spinning universe.